\documentclass{article}
\usepackage[T1]{fontenc}
\usepackage{amstext}
\usepackage{amsbsy}
\usepackage{a4, isolatin1}
\usepackage{epsfig}
\newcommand*\de{\mathrm{d}}
\renewcommand*\epsilon{\varepsilon}
\renewcommand*\phi{\varphi}
\renewcommand*\theta{\vartheta}

\begin{document}
\setlength{\parskip}{2ex}                           
\title{Gravitation in Flat Spacetime}
\author{H. Goenner and M. Leclerc \\ Institut für Theoretische Physik\\ 
Universität Göttingen\\Bunsenstr.9\\Germany}
%\renewcommand{\thefootnote}
%{\fnsymbol{footnote}}
%\footnote{goenner@theorie.physik.uni-goettingen.de} 
\maketitle

%%%%%%%%%%%%%%%%%%%%%%%%%%%%%%%%%%%%%%%%%%%%%%%%%%%%%%%%%%%%%%%%%%%%%%%%%%%%%%%
%%%%%%%%%%%%%%%%%%%%%%%%%%%%%%%%%%%%%%%%%%%%%%%%%%%%%%%%%%%%%%%%%%%%%%%%%%%%%%%

\begin{abstract}
A special-relativistic scalar-vector theory of gravitation is
presented which mimics an important class of solutions of Einstein's
gravitational field equations. The theory includes solutions equivalent to the
Schwarzschild, Kerr, Reissner-Nordström, and Friedman metrics of
general relativity as well as to gravitational waves with parallel
planes. In fact, all the empirical tests until now due to
general relativity can also be explained within this flat spacetime theory.
In order to obtain this result, a new clock hypothesis different from
the one used in special relativity must be introduced. The theory can
be regarded as an example supporting Poincar\'e's conventionality
hypothesis.
\end{abstract}
\pagebreak
%%%%%%%%%%%%%%%%%%%%%%%%%%%%%%%%%%%%%%%%%%%%%%%%%%%%%%%%%%%%%%%%%%%%%%%%%%%%%
%%%%%%%%%%%%%%%%%%%%%%%%%%%%%%%%%%%%%%%%%%%%%%%%%%%%%%%%%%%%%%%%%%%%%%%%%%%%%
\section{Introduction}
In spite of Einstein's extremely successful idea to {\it geometrize} the 
gravitational field with the help of concepts of Riemannian (or rather,
Lorentz-) geometry, notably its {\it curvature}, continued if
unsatisfactory attempts at the formulation of a field theory of gravitation
in {\it flat} spacetime have been made. \footnote{Cf. the discussion
  in \cite{Tor}.} In particular,
R. Feynman believed that the geometric interpretation of gravitation
beyond what is necessary for special relativity, although attractive,
is not essential to physics. \cite{Feyn} 

In the following we suggest a {\it relativistic scalar-vector} theory of
gravitation in {\it flat} spacetime which is exactly equivalent to an 
important subclass of the space of solutions of Einstein's field
equations. This subclass contains those metrics needed for the empirical checks
made until today, i.e. the Schwarzschild-, Kerr-, and
Friedman-Lema\^itre metrics. It also contains classes of
gravitational waves. In Section 2, we describe the simple mapping between the
class of metrics, in Einstein's theory, and the scalar and vector
fields of the flat spacetime theory.

Obviously, in order to retain the well-known effects of gravitational 
red-shift, light-deflection, and perihelion motion in the solar
system, the proposed relations connecting quantities of the mathematical
theory and physical objects (measurement hypotheses) must be changed.
The geodesics of Minkowski space cannot give the correct equations of motion 
for a point particle; in fact, in the new theory the equations of motions do 
not follow from the field equations but must be postulated separately. This is
done in Section 3 where a new clock hypothesis is also proposed. In Section
4, field equations for the scalar and vector fields are written down 
corresponding exactly to Einstein's field equations. In subsequent
sections, we discuss the solution leading to the solar system effects, 
the standard cosmological model, and to Kerr black holes. In a concluding 
section, Poincar\'e's conventionality hypothesis \cite{Poin} will be 
discussed. A possible implication of this approach for a quantum
theory of gravitation is mentioned.

%%%%%%%%%%%%%%%%%%%%%%%%%%%%%%%%%%%%%%%%%%%%%%%%%%%%%%%%%%%%%%%%%%%%%%%%%%%%%%%
%%%%%%%%%%%%%%%%%%%%%%%%%%%%%%%%%%%%%%%%%%%%%%%%%%%%%%%%%%%%%%%%%%%%%%%%%%%%%%%
\section{Introduction of scalar and vector fields through a
  generalized Kerr-Schild metric}
We consider the class of metrics in four-dimensional spacetime
conformal to the Kerr-Schild class:\cite{Kram}
\begin{equation}
g_{ik}=e^{2\sigma}(\eta_{ik}-k_ik_k), (i,k = 0, 1, 2, 3)
\end{equation}
with the Minkowski metric in local inertial coordinates 
$\eta_{ik}=(1,-1,-1,-1)$. The scalar function
$\sigma$ depends on the spacetime coordinates $x^i$. The Kerr-Schild 
vector $k_i(x^k)$ is a null-vector with respect to both metrics $ g_{ik}$
and $\eta_{ik}$:
\begin{equation}
 k^i k^j \eta_{ij} = k^i k^j g_{ij} =  0. 
\end{equation}
In the following, we assume Minkowski space to be the underlying
spacetime. The scalar field $\sigma(x^j)$ and null vector field $k^i(x^j)$ in
flat spacetime are considered as representing the gravitational
field. Thus, in place of the 6 independent (mathematical) degrees of
freedom of general relativity only 4 are retained. At this
point, we warn the reader that we will not aim
at a Maxwellian theory extended by a scalar field, but at a highly
nonlinear scalar-vector theory of gravitation. Although $k^i$
corresponds to a (gravitational) vector potential, no U(1)-gauge group is 
present.

As we remain in flat spacetime, all indices are raised and lowered
with respect to the Minkowski metric, e.g. $k_i = \eta_{ik}k^k$. Thus
for example, differentiation and the rising/lowering of indices commute: 
$\frac{\partial A_k}{\partial x^i}
=A_{k,i}=\eta_{kl}A^l_{,i}=\eta_{kl}\frac{\partial A^l}{\partial x^i}$.

%%%%%%%%%%%%%%%%%%%%%%%%%%%%%%%%%%%%%%%%%%%%%%%%%%%%%%%%%%%%%%%%%%%%%%%%%%%%%%
%%%%%%%%%%%%%%%%%%%%%%%%%%%%%%%%%%%%%%%%%%%%%%%%%%%%%%%%%%%%%%%%%%%%%%%%%%%%%%
\section{Equations of motion and clock hypothesis}
As in Maxwell's theory, field equations and equations of motion will
have to be postulated separately. We begin with the equations of
motion for point particles. Our starting point is the following Lagrangian in
Minkowskian spacetime:
\begin{equation}
L = e^{2\sigma}(u_iu^i - (k_iu^i)^2),
\end{equation}
with the timelike vector of four-velocity $u^i = \frac{\de x^i}{\de \tau}$.
The parameter $\tau$ will be fixed in the following. 
Of course, the Lagrangian (3) corresponds to the one used in general
relativity, $L = g_{ik}u^iu^k$ which leads to the geodesic equations
in curved spacetime with the metric (1). However, in the following, 
we shall forget this \textit{geometric background} and consider expression
(3) as a special relativistic Lagrangian leading to equations of motion in 
Minkowskian spacetime in the presence of gravitational fields $k_i$ and 
$\sigma$. If, in addition, electromagnetic fields are also present, 
the Lagrangian is altered in the usual way: 
\begin{displaymath}
L = \frac{1}{2}e^{2\sigma}(u_iu^i-(k_iu^i)^2)+\frac{e}{m}A_iu^i,
\end{displaymath}
where $A_i$ is the electromagnetic vector potential.
By using the Euler-Lagrange equations known from special relativity,
\begin{displaymath}
\frac{\partial L}{\partial x^m}-\frac{d}{d \tau}\frac{\partial L}{\partial u^m}=0,
\end{displaymath}
we obtain the following equations of motion:
\begin{eqnarray}
(\eta_{im}-k_ik_m)\dot u^i &=& [\sigma_{,m}\eta_{ik}
-2\sigma_{,k}\eta_{im}]u^iu^k - [\sigma_{,m}k_ik_k-
2\sigma_{,k}k_ik_m]u^iu^k\nonumber \\& & -\frac{1}{2}
[(k_ik_k)_{,m}-2(k_ik_m)_{,k}]u^iu^k. 
\end{eqnarray}
Just as in special relativity, these 4 equations are not independent. 
Multiplying with $u^m$, one easily finds a first integral: 
$L=\epsilon = const$. To fix the parameter $\epsilon$, we look at
$u_iu^i = e^{-2\sigma}\epsilon +(k_iu^i)^2$, which, in the absence
of any gravitational field, reduces to $u_iu^i = \epsilon$. Thus we
conclude from special relativity (free motion in Minkowski spacetime) 
that $\epsilon$ should be set to unity (by putting $c = 1$) for 
the motion of massive particles. When describing null curves, $u_i$ is
to be understood
as the wave-vector of the electromagnetic radiation field,
(cf. \cite{Lan}, p. 154) and we should have $\epsilon = 0$. By this, the 
parameter $\tau$ of the curve is fixed modulo multiplication with a
constant factor in the case $\epsilon = 0$. 

Thus, we have three independent equations of motion for the four
independent functions (three components of $k_i$ and the scalar 
function $\sigma$) describing the gravitational field. In $\epsilon$
we recognize  a type of energy (or mass) parameter. 

Now, a first digression from a typical special relativistic field
theory is needed. In special relativity theory, the clock hypothesis
relates Minkowskian proper time with what is measured by a clock in an
arbitrary location, and in an arbitrary state of motion. Obviously, 
gravitational red-shift cannot be obtained in this way. 

To give it a physical meaning, we imagine ``the observer'' or ``clock''
to coincide with a radiating atom etc; consequently, in the equations
of motion we will have to set $\epsilon = 1$. We define ``proper'' time as
the parameter $\tau$ of the path of
an observer at rest with respect to his coordinate frame, i.e an observer
on a curve with $u^{\alpha} = 0$ for $\alpha = 1,\ 2,\ 3$. Next, we
identify this modified ``proper'' time with the physical time, the time
measured by a clock moving along this curve. 

From $L = \epsilon =  e^{2\sigma}(u_iu^i-(k_iu^i)^2)$, 
with $u^{\alpha} = 0$, the
relation between ``proper'' time $\tau$ and the time coordinate $t$ is
easily found to be ({\it new} clock hypothesis):
\begin{equation}
u_0 = \frac{\de t }{\de \tau} = \frac{e^{-\sigma}}{\sqrt{1-k_0^{\ 2}}}.
\end{equation}
Of course, as in any special-relativistic theory, the relation between
``proper'' time and the coordinate time of a second observer, moving 
uniformly and linearly with respect to the first one, is given by a 
Lorentz transform of the time coordinate. \footnote{By this definition
we are not in a position to define ``proper'' time for a general observer.
However, this lack is easily mended if we define ``proper'' time through the 
Kerr-Schild-metric (1) without taking recourse to its geometrical meaning.}
In the following, we strictly adhere to inertial systems only, i.e to the 
Minkowski metric in locally inertial (cartesian) coordinates.

In equation (5) we recognize a condition on the gravitational field
$k_i$, namely $k_0^{\ 2} < 1$. We shall return to it in section 5.
Comparing two clocks at different positions at the same coordinate time $t$, we
are led to gravitational redshifts as they are known from general 
relativity: 
\begin{equation}
z = \frac{\omega_1}{\omega_2}-1 = \frac{\de \tau_2}{\de \tau_1}-1=
 \frac{e^{\sigma(t,\boldsymbol{x_2})}
\sqrt{1-k_0(t,\boldsymbol{x_2})^2}}
{e^{\sigma(t,\boldsymbol{x_1})}
\sqrt{1-k_0(t,\boldsymbol{x_1})^2}}\ \ -1.
\end{equation}

The results of this section will be used in the discussion of some 
concrete examples in section 5. 

%%%%%%%%%%%%%%%%%%%%%%%%%%%%%%%%%%%%%%%%%%%%%%%%%%%%%%%%%%%%%%%%%%%%%%%%%%%%%%
%%%%%%%%%%%%%%%%%%%%%%%%%%%%%%%%%%%%%%%%%%%%%%%%%%%%%%%%%%%%%%%%%%%%%%%%%%%%%%
\section{Field equations of the scalar-vector theory in flat spacetime}
As we want to establish a special relativistic theory of
gravity, we cannot use a geometric concept as the curvature
of spacetime in order to obtain the needed field
equations. Nevertheless, we try to stay as close as possible to the 
Einstein equations in order to retain solutions as Schwarzschild's or 
Friedman's. As we shall see in section 5, the 
Schwarzschild solution is given by $\sigma = 0$, and the vector field 
$k^i = \sqrt{\frac{a}{r}} (1,\ -\frac{\boldsymbol x}{r})$ with the
constant a. 
The first idea would be to take the action-functional from
general relativity, i.e. $S = \int{\sqrt{-g}R\  \de^4x}$, with the
Ricci scalar $R$ of the metric (1) (in cartesian coordinates 
$(t,x,y,z)$) expressed in terms of $k^i$ and $\sigma$, and to carry 
out the variation not with respect to the metric $\eta_{ij}$, but with 
respect to the gravitational fields $k^i$ and $\sigma$. This procedure
leads to a vector and a scalar equation. However, the following two possible
processes cannot commute. I: As a first step, insert (1) into the
Lagreangian, then vary with regard to the fields $\sigma$ and $k^i$, and
II: As a first step vary with respect to the Kerr-Schild-metric, then
insert (1) into the resulting ten equations.

A straightforward calculation shows that the Schwarzschild solution
is {\it not} a unique solution of the field equations following from 
procedure I. Hence we base our special relativistic theory of gravitation on
procedure II, i.e. on the Einstein vacuum field equations expressed as
field equations for $\sigma, k^i$ in Minkowski spacetime:
 \begin{eqnarray}
\Gamma_{ik}&=&\frac{1}{2}[-(k^lk_k)_{,i,l}-(k^lk_i)_{,k,l}
+(k^lk^m)_{,l}(k_ik_k)_{,m}-(k^mk_i)_{,l}(k^lk_k)_{,m}
+\nonumber \\ & &
+(k_kk_i)_{,m}^{,m}+k^lk^m(k_kk_i)_{,l,m}+k_ik_kk_{l,a}k^{l,a}
+k^mk^a(k_ik_k)k_{l,a}k^l_{,m}]+\nonumber \\ & &
+k_ik_k(k^lk^m\sigma_{,m})_{,l}+k_ik_k\sigma^{,m}_{,m}+
2k_ik_k\sigma_{,m}\sigma^{,m}+2k_ik_kk^lk^m\sigma_{,m}
\sigma_{,l}-\nonumber \\ & &
-2\sigma_{,i,k}+
2\sigma_{,i}\sigma_{,k}+ (k_ik_k)_{,l}
\sigma^{,l}+k^lk^b(k_ik_k)_{,l}\sigma_{,b}-(k^lk_i)_{,k}
\sigma_{,l}-\nonumber\\& &-(k^lk_k)_{,i}\sigma_{,l}+
\eta_{ik}[-(k^lk^m\sigma_{,m})_{,l}-\sigma_{,m}^{,m}-
2\sigma_{,m}\sigma^{,m}-2k^lk^m\sigma_{,m}\sigma_{,l}]\nonumber\\
&=&0.
\end{eqnarray}
After having written down the field equations we may forget the connection to 
general relativity. We shall use these equations in the geometric background 
of Minkowski space, only. In the form given, the field equations are 
covariant only with respect to the Poincar\'e group, i.e. under 
transformations of the form
\begin{displaymath}
x^{i'} = \Lambda^i_{\ k}x^k+a^i,
\end{displaymath}
with $\det(\Lambda^i_{\ k})= \pm 1$.

Now, a second new postulate for the coupling of the
gravitational field $\sigma, k^i$ to its matter sources is
needed. Matter will be described by its energy-stress tensor, coupled to the
gravitational fields by the coupling constant $\kappa$ known from
general relativity. The energy-stress tensor is found as in any
other special-relativistic field theory by symmetrization of 
the canonical tensor \cite{Das} 
\begin{equation}
T^{\ k}_i = q^l_{,i}\ \frac{\partial \Lambda}
{\partial q^l_{,k}}
 - \delta^k_i \Lambda.
\end{equation}

Starting from the \textit{mixed} tensor
$T_i^{\ k}$ defined in (8), using the following definitions:
\begin{displaymath}
\tilde T_{ik} = g_{lk}T_i^{\ l},\ \ 
T_{ik}= \eta_{lk} T_i^{\ l},\ \ 
\tilde T = \delta^i_k T_i^{\ k}
=T, 
\end{displaymath}
and via $\Gamma_{ik} = \kappa~(\tilde T_{ik} - 1/2~g_{ik} \tilde
T^l_l)$, and the metric (1), we arrive at the equations:
\begin{equation}
\Gamma_{ik}= \kappa e^{2\sigma}[T_{ik}-\frac{1}{2}
(k_kk^mT_{im}+k_ik^mT_{km})-\frac{1}{2}(\eta_{ik}-k_ik_k)T].  
\end{equation}
The Lagrangian density $\Lambda$ should describe the whole material system
in interaction with the gravitational fields $\sigma, k^i$, i.e. in the
regular case it will also contain the fields $\sigma, k^i$. We shall
give examples in section 5.

In an
alternative approach, we tried to find a Lagrangian density leading to
vector and scalar equations (again by variation with respect to the
fields $\sigma, k^i$) which are equivalent to the Einstein vacuum
field equations. Only partial success has been reached: a Lagrangian
has been found the variation of which with respect to $\sigma, k^i$
leads to the equations:
\begin{equation} \Gamma_{ij}k^j = 0 , ~  \Gamma_{ij}\eta^{ij} = 0. 
\end{equation}

This Lagrangian is given in Appendix 1. However, it is not good enough. 
It can be shown that no combination of
any two of the following equations  $R=0$, $R_{ik}k^k=0$, 
$R^{ik}\sigma_{,k}=0$, or $R^{ik}\sigma_{,i,k}=0$ is sufficient to fulfill 
Birkhoffs theorem, i.e. to guarantee uniqueness of the spherically symmetric
vacuum solution. 

%%%%%%%%%%%%%%%%%%%%%%%%%%%%%%%%%%%%%%%%%%%%%%%%%%%%%%%%%%%%%%%%%%%%%%%%%%%%%
%%%%%%%%%%%%%%%%%%%%%%%%%%%%%%%%%%%%%%%%%%%%%%%%%%%%%%%%%%%%%%%%%%%%%%%%%%%%%
\section{Some solutions of physical relevance}
We are now in a position to consider particular solutions of the field
equations of the scalar-vector theory in flat spacetime, 
and to compare the results with those from general relativity.
In this section, the scalar field is assumed to vanish: $\sigma = 0$
while $k^i(x^k) \neq 0$.

\subsection*{Spherically symmetric solutions} 

i. {\it The vacuum case.}\\

The general spherically symmetric
null-vector field in cartesian coordinates is given by
 \begin{displaymath}
k_i = f(r,t)(1,\ \pm \frac{\boldsymbol x}{r}).
\end{displaymath}
Introducing this expression into the vacuum field equations (7), we are led to
\begin{equation}
k_i = \pm \sqrt{\frac{a}{r}}\ (1,\ \pm\frac{\boldsymbol x}{r}),
\end{equation}
with constant $a $. The global sign of the vector field is
irrrelevant, because all physical quantities are obtained from the 
Lagrangian (3). 

In general relativity, to the Lagrangian (12) a line-element 
\begin{displaymath}
\de s^2 = \de t^2 - \de r^2 -r^2 \de \Omega^2 - \frac{a }{r}(\de r \mp
\de t)^2,
\end{displaymath} 
corresponds which is the Eddington-Finkelstein form of the 
Schwarzschild metric.\cite{Pap} It may be brought into the classical 
Schwarzschild form 
\begin{displaymath}
\de s^2 = (1- \frac{a }{r})\de \tau^2 - (1-\frac{a }{r})^{-1}\de 
r^2 -r^2 \de \Omega^2
\end{displaymath} 
by the coordinate transformation
\begin{displaymath}
\tau = t \pm a  \ln (\frac{r}{a } - 1).
\end{displaymath}

From this solution it is noted, that the condition $k_0^{\ 2} < 1$ derived
in section 3 corresponds to $g_{00}>0$ in general relativity. If this 
condition is violated, the corresponding frame of reference cannot be realized
by real bodies. \cite{Lan} \footnote{We also know from general relativity 
that no geodesic passes through the Schwarzschild radius $a $ from
inside, and a massive particle outside this radius can only reach 
the Schwarzschild radius after an infinite time.} In our flat spacetime
theory, we simply restrict the validity of the solutions of 
the field equations to regions where $k_0^{\ 2}< 1$, i.e. r > a.

In spherical coordinates, the Lagrangian  for the
equations of motion is given by
\begin{equation}
L = \dot{t}^2 -r^2(\sin^2 \vartheta\dot{\varphi}^2
+\dot{\vartheta}^2)-\dot{r}^2 -\frac{a }{r}(\dot{t}\pm\dot{r})^2,
\end{equation}
where a dot means differentiation with respect to the parameter along the
path. It can easily be seen that the change of sign of the spatial 
coordinates of $k^i$ corresponds to the substitution $t \rightarrow -t$.
In the following, we are using the $+$ sign in (11).
The Euler-Lagrange equation for $\theta$ has the special solution $\theta =
\frac{\pi}{2}$; the cyclic coordinates $t$ and $\phi$ lead to 
the integrals
\begin{eqnarray*} 
d&=&\dot{t} - (\dot{t}+\dot{r})a /r \\ 
l&=&r^2\dot{\varphi}.
\end{eqnarray*}
Another integral is, as we saw in section 3, the Lagrangian itself, 
$L=\epsilon$, which gives  a differential equation for the 
radial coordinate
\begin{displaymath}
\dot{r}^2=d^2-(\varepsilon+\frac{l^2}{r^2})(1-\frac{a }{r}).
\end{displaymath}
or, introducing the variable $\rho = 1/r$ and 
\begin{displaymath}
\dot{r}=\frac{\de r}{\de \rho}\dot{\rho}
       =-\frac{1}{\rho^2}\frac{\de\rho}{\de\varphi}
\dot{\varphi}
      =-\frac{1}{\rho^2}\frac{\de\rho}{\de\varphi}
\frac{l}{r^2}
= -l \frac{\de \rho}{\de \varphi},
\end{displaymath}
we come to the following equation:
\begin{equation}
(\frac{\de \rho}{\de \phi})^2 = \frac{d^2 - \epsilon}{l^2}
+ \frac{a \epsilon}{l^2} \rho - \rho^2 + a  \rho^3.
\end{equation}
We see that the substitution $t \rightarrow -t$ has no effect
on this equation and its solutions. Comparing equation (13) with
the results of general relativity, we see, by setting 
$a  = 2\gamma M$, with the newtonian coupling constant $\gamma$
and the mass $M$ of a central body, that we found exactly the
same differential equation as for the geodesics of the 
Schwarzschild metric \footnote{Cf. many textbooks on general
  relativity, e.g. \cite{Goe1}, equation (10.23).}, for both
$\epsilon =0$ and $\epsilon =1$. Hence, without any further 
calculation, we know that the solution of (13) for massive
particles ($\epsilon = 1$) will lead to the well-known
\textit{perihelion motion} which means that for every rotation around 
the center of symmetry, the position of the perihel changes 
by $\Delta \phi \approx \pi \frac{3a ^2}{2l^2}$. By putting $\epsilon
= 0$, a second well-known result, the deflection of lightrays 
grazing the rim of the sun, will be obtained. Just as in general
relativity, we find $\Delta \phi = 2 \frac{a }{D}=\frac{4\gamma
  M}{D}$, with $D$ the radius of the sun. How to obtain these results
from (13) can be found in any standard textbook on general relativity. 

A further experimental result, gravitational redshift, arises if
simultaneous readings of two clocks at different
altitudes are made. Using equation (6), 
we have 
\begin{displaymath}
z = \frac{\omega_1}{\omega_2}-1 = \sqrt{\frac{1-\frac{a }
{r_2}}{1-\frac{a }{r_1}}},
\end{displaymath}
which is in total agreement with general
relativity. \footnote{\cite{Goe1}, equation (10.75). This means that a 
clock at a higher altitude runs faster than one at sea level.}

ii. {\it Electromagnetic fields} \\

We now consider spherically symmetric solutions with both gravitational
and {\it electromagnetic} fields. We start from the following matter
Lagrangian density for the electromagnetic field:
\begin{displaymath}
\Lambda= -\frac{1}{4}F^{lm}F_{lm}+
\frac{1}{2}F^{lm}F_{lk}k_mk^k-
e^{4\sigma} A_mj^m,
\end{displaymath}
where the electromagnetic potential $A_i$ is defined in the 
usual way by $F_{ik}= A_{k,i}-A_{i,k}$.
It can easily be shown that variation with respect to $A_i$ leads to
\begin{equation}
F^{ik}_{\ ,k}-(k^ik_kF^{kl})_{,l}-(k^kk_lF^{il})_{,k}=-e^
{4\sigma}j^i,
\end{equation}
which reduces to the usual Maxwell equations in the case 
of vanishing gravitational fields. We have to solve the field equation
(9) by using (8) with the above Lagrangian density. By the definition
of the potential $A_i$, the first group of Maxwell equations remains unchanged.
Therefore, we still have gauge invariance with respect to 
transformations of the form $\tilde A_i = A_i + \lambda_{,i}$. 
Thus, in the spherically symmetric ansatz 
\begin{eqnarray}
A_i &=& (\phi(r),\  \psi(r)\frac{\boldsymbol{x}}{r})\\
k_i &=& f(r)\ (1,\  \frac{\boldsymbol x}{r})
\end{eqnarray}
$\psi$ can be transformed to zero. Subtitution into (14) leads to 
the solution for the electromagnetic potential
 \begin{displaymath}
A_i = (\phi(r),\ 0,\ 0,\ 0),
\end{displaymath}
where 
\begin{displaymath}
\phi(r) = \frac{e}{r},
\end{displaymath}
$e$ being a constant of integration.
Solving equations (9) for the gravitational field, one finds after
some work
\begin{equation}
k_i = \sqrt{-\frac{\kappa e^2}{2r^2} + \frac{a }{r}}\ \  (1,\ 
\frac{\boldsymbol x}{r}).
\end{equation}
By putting either $e = 0$ or $a =0$, and by comparing with
(11) or with the Coulomb potential, respectively, the two constants of 
integration are identified as charge of the central particle $e$ and the 
Schwarzschild radius $a $. We will not discuss this solution, as it 
corresponds entirely to the Reissner-Nordström metric of general
relativity. \footnote{Cf. \cite{Pap} Apart from the Bertotti-Robinson
solution this is  the only spherically symmetric solution of the
combined Einstein-Maxwell equations.}   

\subsection*{Further solutions}
The well-known Kerr metric descibing a rotating black hole \footnote{For 
information about Kerr-Schild metrics cf. \cite{Kram}~\cite{Sop}~\cite{Bis}} can now be found as a vacuum solution of
equation (7) in the form 
\begin{eqnarray*}
k_0 &=& \sqrt{\frac{Rr^3}{r^4+a^2z^2}}\\
k_x &=& \sqrt{\frac{Rr^3}{r^4+a^2z^2}}\ \ [\frac{r}{r^2+a^2}\ x 
+\frac{a}{r^2+a^2}\ y]\\
k_y &=& \sqrt{\frac{Rr^3}{r^4+a^2z^2}}\ \ [\frac{r}{r^2+a^2}\ y 
-\frac{a}{r^2+a^2}\ x]\\
k_z &=& \sqrt{\frac{Rr^3}{r^4+a^2z^2}}\ \ \frac{z}{r},
\end{eqnarray*}
with the Schwarzschild radius R and angular momentum parameter a.

Wave solutions can be obtained if the condition $(k^ik^k)_{,i}=0$ is
satisfied. In the first order approximation of an expansion in $k^i$, i.e. if 
only terms quadratic in $k^i$ are retained, equation (7) reduces to 
$(k_ik_k)_{,m}^{,m}=0$, the wave equation in Minkowski space. These fields 
correspond to gravitational waves described in general relativity by a 
linearized metric $g_{ik}=\eta_{ik}+h_{ik}$. (Cf. \cite{Lan}, §107, p. 411.)

Exact wave solutions follow if the ansatz 
\begin{displaymath}
k_i = \sqrt{H(t-x, y, z)}\ (1,\ -1,\ 0,\ 0)
\end{displaymath}
is inserted into the vacuum field equations of the scalar-vector theory.
Equation (7) reduces to $H_{,y,y}+H_{,z,z}=0$, the differential equation
for the so called pp-waves. In general relativity, they correspond to a metric
$\de s^2 = 2 \de u \de v - 2 H(u,y,z)\de u^2-\de y^2 - \de z^2$, with
$t = \frac{1}{\sqrt 2}(v+u)$ and $x = \frac{1}{\sqrt 2}(v-u)$.
(\cite{Lan}, §109, p. 419.)

As an example for solutions involving both fields $\sigma \neq 0$ and 
$k^i \neq 0$,  the interior Schwarzschild solution is listed in 
appendix 2. \footnote{The
  particular form of $\sigma$ and $k^i$ can be obtained from
 \cite{Dad}.}

%%%%%%%%%%%%%%%%%%%%%%%%%%%%%%%%%%%%%%%%%%%%%%%%%%%%%%%%%%%%%%%%%%%%%%%
%%%%%%%%%%%%%%%%%%%%%%%%%%%%%%%%%%%%%%%%%%%%%%%%%%%%%%%%%%%%%%%%%%%%%%
\section{Cosmological solutions}
We now turn to solutions with $k_i=0,\ \  \sigma = \sigma(r,t)
\neq 0$. The remaining field equations are:
\begin{displaymath}
-2\sigma_{,i,k}+2\sigma_{,i}\sigma_{,k}-\eta_{ik}[\sigma_{,m}
^{m}
+2\sigma_{,m} \sigma^{,m}]=\kappa e^{2\sigma}
[T_{ik}-\frac{1}{2}
\eta_{ik}T].
\end{displaymath}
The matter distribution is described by the energy-stress tensor
of a perfect fluid, chosen to be
\begin{displaymath}
T^{ik}= e^{2\sigma}(p+\mu) u^iu^k - p\eta^{ik}.
\end{displaymath}
This tensor reduces to the usual diagonal special relativistic tensor 
in the comoving frame, where $u^i = e^{-\sigma}(1,\ 0,\ 0,\ 0)$ (for
$k_i=0$).
The four remaining independent equations are the following:
 \begin{eqnarray}
e^{-2\sigma}[\sigma''-3\ddot{\sigma}+2\sigma'^2+2\sigma'/r]
&=&\kappa[e^{2\sigma}(\mu+p)u_0^{\ 2}-\frac{1}{2}(\mu-p)]
\nonumber \\
e^{-2\sigma}[-5\sigma''+3\ddot{\sigma}-10\sigma'/r-4\sigma'^2
+6\dot{\sigma}^2]&=&\kappa[e^{2\sigma}(\mu+p)\boldsymbol{u}^2
+\frac{3}{2}(\mu -p)]\nonumber \\
e^{-2\sigma}[-2\sigma''+2\sigma'/r+2\sigma'^2]&=&\kappa e^{2
\sigma}[(\mu+p)u_xu_y]\frac{r^2}{xy} \nonumber\\
e^{-2\sigma}[-2\dot{\sigma'}+2\dot{\sigma}\sigma']&=&\kappa
e^{2\sigma}[(\mu+p)u_0u_x] \frac{r}{x},
\end{eqnarray}
where $\boldsymbol{u}^2 =(u^x)^2+(u^y)^2+(u^z)^2$. The dot and
strike stand for differentiation with respect to the time
and radial coordinates, respectively. For simplicity, we look
for solutions in a localy comoving frame, i.e. a frame in which the
matter distribution is described by $u^i = e^{-\sigma}\delta_0^i$.

In addition, a barotropic equation of state  $p = b \mu$ with constant
$b$ is assumed. For dust matter, i.e. for  $b = 0, \mu$ finite,
equations (18) lead to the unique solution:
\begin{eqnarray}
\kappa p&=&0 \nonumber \\
\kappa \mu &=& 12a^2(t-t_0)^{-6}\nonumber\\
e^{2\sigma}&=& a^{-2} (t-t_0)^4.
\end{eqnarray}
For radiation, i.e. $\mu = 3p$, the only solution reads as:
\begin{eqnarray}
\kappa \mu &=& 3\kappa p\  =\ 3 a^2
(t-t_0)^{-4}\nonumber\\
e^{2\sigma} &=& a^{-2} (t-t_0)^2.
\end{eqnarray} 
A unique, inhomogeneous solution (depending on the radial coordinate) is given
by 
\begin{eqnarray}
\kappa \mu &=& - \kappa p\ =\ 12\lambda c + 3 a^2\ =\ const 
\nonumber \\
e^{2\sigma} &=& [\lambda(r^2-t^2)+at+c]^{-2}.
\end{eqnarray}
Both for $p = 0$ or $\mu = 3p$ there exist further solutions , since
we can rewin all conformally flat solutions of Einstein´s equations
from (18), in particular, the Friedman-Robertson-Walker solutions. 
We did restrict ourselves to matter distributions with 
$u^i = \delta_0^i e^{-\sigma}$.\footnote{This requires either that there 
is an inertial frame of reference in which the matter is at rest,
or that we consider only local properties of the solutions, since
by a Lorentz boost, we can always find a locally comoving frame.} 

Further solutions can be obtained through a more general ansatz with
non-aligned $u^i$ or, vice versa, by starting with the 
Friedman-Robertson-Walker form of the metric
in general relativity in the conformally flat form, and by
performing the same transformation on $u^i$.\footnote{More on the two 
equivalent representations of cosmological models within the framework of 
Einstein´s theory, namely the conformally flat and the spatially 
homogenous and isotropic Robertson-Walker form, can be found in
\cite{Que}, \cite{Kea}.} 

Using equation (5), we introduce the physical quantity $\tau$ into
our solutions (19) and (20) to obtain $\kappa\mu = \frac{4}{3}(\tau-\tau_0)^{-2}
$ and $\kappa\mu = \frac{3}{4}(\tau-\tau_0)^{-2}$, respectively. Turning to 
solution (19), i.e. to spacetime filled with dust matter, and a \textit{fixed
amount of dust} of constant mass $M$ contained in a sphere of radius $R$,
we obtain from $M = \frac{4\pi}{3} \mu R^3$ the expression $R(t)^3=\frac{3M}{4\pi}\mu(t)^{-1}$,
or, with (19) and the variable $\tau$,
$R(\tau)=(\frac{9M}{16\pi a^2})^{1/3}\ (\tau-\tau_0)^{2/3}$.
Since $\tau$ and $M$ are physical quantities, we have found a physical
distance quantity $R(\tau)$. We can generalize this to the statement
that any physical distance between two points in spacetime with constant
coordinate distance depends on the physical time coordinate through
\begin{displaymath}
l(\tau) \sim (\tau-\tau_0)^{2/3}.
\end{displaymath} 
This is in fact the result known from general relativity, where
the \textit{world-radius} $S(\tau)$ shows the same time dependence
in the dust-matter cosmos ($p=0$) with space curvature $k=0$.\footnote{Locally,
(i.e. for small $S$) this is valid also in the open and closed models
($k=\pm1$) with $p=0$. (Cf. \cite{Ray}, section 1.2)} 

For the solution (20) corresponding to radiation as a material source,
we define length-intervals by light propagation. It is easily seen that for 
radial light rays, the following solutions of our equations of motion (c.f. 
section 3) are obtained:
$r=\pm (t-t_0)$, i.e. $r \sim (\tau-\tau_0)^\frac{1}{2}$.
We can interpret $r$ as a measurable quantity representing
the radius of an expanding wave-front, for instance. Generalizing it 
to a physically meaningful distance $l(\tau)$, we obtain 
\begin{displaymath}
l(\tau) \sim (\tau -\tau_0)^{1/2},
\end{displaymath}
which again corresponds to the results known from the Friedman solutions
with $\mu = 3p$. (\cite{Ray}, section 16.2.)

Now cosmological redshift will be briefly discussed. We cannot use
equation (6), because it refers to the relation between 
two clocks read simultaneously at different places. In the
cosmological context we have to compare the frequency $\omega_0$ of 
a signal emitted at $(r,t)$ with the frequency $\omega_1$ of the same
signal after it reached the observer at $(r_1, t_1)$. From 
$z= \frac{\omega_0}{\omega_1}-1=\frac{\de \tau_1}{\de \tau}-1
= e^{\sigma(t_1,r_1)-\sigma(t,r)}$, 
we obtain for the  cosmological redshift of a source at $(r_1, t_1)$:
$z= (\frac{t_1}{t})^2 -1$ for the solution (21) and $z=\frac{t_1}{t}-1$
for (20). If again we assume radial light propagation, we obtain from 
$L = e^{2\sigma}(\dot t^2 - \dot r^2)=0$ (the dot refers to the 
parameter of the curve, which is not identical with ``proper'' time $\tau$
defined on curves with $L = \epsilon = 1$): $r(t)=r_1-(t-t_0)$, where 
$(r_1,t_1)$ are
the coordinates of signal absorption. By this, we can 
express the redshift as a function of the coordinate distance $r-r_1$ 
between the source ($r$) and the observer ($r_1$). However, it is more usefull
to express the redshift by physical quantities. This is 
not hard to to since we know the relation between coordinate time $t$ 
and the physical quantity $\tau$. 
The results are $z = (\frac{\tau_1}{\tau})^{2/3}-1$
and $z = (\frac{\tau_1}{\tau})^{1/2}-1$ for
the solutions (19) and (20), respectively. Again, these results
are in complete agreement with general
relativity. (\cite{Goe2}, sections 2.4.2 and 3.2.1, or other
textbooks.) The same results are derived 
by considering the time dependence of the 
wavelength (as a measurable quantity), $\lambda \sim \tau^{3/2}$ and
$\lambda \sim \tau^{1/2}$ for the solutions (19) and (20) resp., and
upon using $z=\frac{\lambda(\tau_1)}{\lambda(\tau)}-1$. 

The solution (21) describes a physical system with b = -1, i.e. the
equation of state $\mu = -p$. Such an equation of state arises
in inflationary cosmological models.

Repeating the arguments used for the case $p=0$ applied to a sphere 
containing dust of a fixed mass, since $p$ and $\mu$ are time-independent, we 
find that the physical distance is identical to the coordinate distance, i.e. 
in our comoving frame the physical distance between any \textit{dust particles}
does not change in time. Thus, (21) represents a static cosmos. By the 
same argument as before ($z= \frac{\lambda(\tau_1)}{\lambda(\tau)}-1$), we 
find that there is no redshift occurring.

If no equation of state is prescribed, further solutions of the field equations
(9) can be obtained. We list one of them without attempting to provide an 
interpretation:
\begin{eqnarray*}
\kappa \mu &=& 12 \lambda b\nonumber \\
\kappa  p  &=& 8 \lambda^2 r^2 - 16 \lambda b\nonumber \\
e^{2\sigma}&=& (b + \lambda r^2)^{-2},
\end{eqnarray*}
with $\lambda, b=const$, which could describe a positive energy distribution
in a sphere of radius $r_0 = 2b/\lambda$. At $r_0$, the pressure $p$ is
vanishing, and we can complete the solution by joining to it the spherically 
symmetric vacuum solution. 

%%%%%%%%%%%%%%%%%%%%%%%%%%%%%%%%%%%%%%%%%%%%%%%%%%%%%%%%%%%%%%%%%%%%%%%%%%%%
%%%%%%%%%%%%%%%%%%%%%%%%%%%%%%%%%%%%%%%%%%%%%%%%%%%%%%%%%%%%%%%%%%%%%%%%%%%
\section{Discussion}
The special-relativistic scalar-vector theory of gravitation presented here 
forms an example for Poincar\'e's conventionality hypothesis stating that  
empirical data always can be explained by different theories based on
different hypotheses. Without using all the geometrical concepts of
Einstein's theory of gravitation, the scalar-vector theory predicts
the well-known effects in the solar system, of the standard
cosmological model, of black holes; it also allows for gravitational 
waves. 

We do not suggest the new theory as a serious competitor for general 
relativity: as a field theory in Minkowski space it is rather complicated. 
Nobody would have thought of the particular field equations suggested
before the event of Einstein's theory. Also, a Lagrangian still is to
be found from which, after variation with respect to $\sigma, k^i$, field 
equations allowing for Birkhoff's theorem will follow. At present, the field 
equations equivalent to Einstein's are written down ad hoc. \footnote{A method
for deriving them formally from a variational principle by use of Lagrangian 
multiplicators will be discussed elsewhere.}. A Lagrangian 
approach already available
leads to further spherically symmetric vacuum field equations beyond 
Schwarzschild's. (Cf. appendix 1.) The equations of motion do not follow from 
the field equations, but must be postulated separately as in Maxwell's theory.
Moreover, cartesian coordinates in Minkowski spacetime have no
physical meaning. New measurable quantities for time and distance have
been introduced. 

On the other hand, a scalar-vector field theory in flat
spacetime should be more amenable to the standard recipees for quantization.
Hence, upon assuming that progress can be made towards both a manageable 
Lagrangian for the field equations equivalent to (9), or (7), and  the 
well-posedness of the Cauchy initial value problem, there might be a 
possibility for a viable theory of quantum gravitation equivalent to part of 
the quantization of the full Einstein theory. It may be, however, that
the theory is as unrenormalizable as general relativity.

As to Poincar\'e's hypothesis: by adding a principle of simplicity to
it, in most cases a decision can be made as to which theory is
preferable - although ``simplicity'' itself may not always be
unambiguously defined. One conclusion of this paper is that in an
important subclass of solutions of Einstein's equations, a second
flat metric appears in a way leading beyond the equivalence principle
to a relativistic scalar-vector theory of gravitation in Minkowski space.

%%%%%%%%%%%%%%%%%%%%%%%%%%%%%%%%%%%%%%%%%%%%%%%%%%%%%%%%%%%%%%%%%%%%%%%%%%%%
%%%%%%%%%%%%%%%%%%%%%%%%%%%%%%%%%%%%%%%%%%%%%%%%%%%%%%%%%%%%%%%%%%%%%%%%%%%

%\footnotesize
\section*{Appendix 1}

In this appendix, we give a  Lagrangian, the variation of which leads
to the equations
\begin{displaymath}
\Gamma_{ij}k^j=0, \ \ \Gamma_{ij}\eta^{ij}=0.
\end{displaymath}

Consider the following Lagrangian, $\lambda$ beeing a lagrangian
multiplier:
\begin{eqnarray*}
\Lambda &=& e^{2\sigma}[-3\sigma_{,i}^{,i}-3\sigma_{,i,k}k^ik^k+\frac{1}{2}
(k^ik^k)_{,i,k}+\frac{1}{2}(k^kk^k_{,l}k^ik_{k,i})] + \\
 && + e^{2\sigma}k_kk^k[k_{i,m}k^{i,m}-k_{m,i}k^{i,m}+6\sigma_{,i,k}k^ik^k
+4k^l_{,i}k^i\sigma_{,l}+\\ && +2k^l_{,l}k^m\sigma_{,m}+2\sigma^{,m}_{,m}
+4\sigma_{,m}\sigma^{,m}+\frac{1}{2}k^lk^mk_{i,l}k^i_{,m}]+
\lambda(k_ik^i)^2.
\end{eqnarray*}

Carrying out the variation with respect to $\lambda$, $k_i$ and 
$\sigma$, we come to the following equations:

\begin{displaymath}
k_ik^i=0,
\end{displaymath}

\begin{eqnarray*}
-\frac{1}{2}(k^lk^ik_{k,i})_{,l}+\frac{1}{2}k^lk^i_{,l}k_{i,k}-
2\sigma_{,i,k}k^i+2\sigma_{,i}\sigma_{,k}k^i -k^lk_{k,i}\sigma_{,l}k^i +&&\\
+[-\frac{1}{2}k_{i,m}k^{i,m}+\frac{1}{2}k_{m,i}k^{i,m}-\frac{1}{2}k^lk^mk_{i,l}
k^i_{,m}-k^l_{,i}k^i\sigma_{,l}-&&\\ -
(k^lk^m\sigma_{,m})_{,l}-\sigma^{,m}_{,m}-2\sigma_{,m}\sigma^{,m}-2k^lk^m
\sigma_{,m}\sigma_{,l}]k_k&=&0,
\end{eqnarray*}

and 

\begin{displaymath}
-(k^ik^k)_{,i,k}+\frac{1}{2}(k^lk^k_{,l})(k^ik_{k,i})-6(\sigma_{,i}^{,i}
+\sigma_{,i}\sigma^{,i})-6k^ik^k(\sigma_{,i,k}+\sigma_{,i}\sigma_{,k})-6
(k^ik^k)_{,i}\sigma_{,k}=0.
\end{displaymath}

The second and third equations indeed correspond
to $\Gamma_{ik}k^i = 0$ and $\eta^{ik}\Gamma_{ik}=0$,
the first beeing just the null-vector condition to the
vector field.

Making the same spherically-symmetric ansatz as in section 
5, we find the general solution (for $\sigma=0$) in the 
form
\begin{displaymath}
k^i=\sqrt{\frac{h(t+r)}{r}}\ (1,\ -\frac{\boldsymbol x}{r}),
\end{displaymath}
the spherically-symmetric vacuum solution thus beeing determined
only up to a  function of $t+r$. Even for these equations for which
Birkhoff's theorem is not satisfied, the Lagrangian is far from 
beeing simple. 

If $\sqrt{-g}R$ is taken as a Lagrangian, the equations found 
after the variation with respect to $\sigma$ and $k_i$ even 
do not have $k^i= \sqrt{\frac{R}{r}}\ (1,\ -\frac{\boldsymbol{x}}{r})$ as
a solution. A possible solution of these equations ($\sigma =0$) is:
$k^i= (1,\ -\frac{\boldsymbol{x}}{r})$, which in spherical coordinates
corresponds to the constant vector $k^i = (1,\ -1,\ 0,\ 0)$.

%%%%%%%%%%%%%%%%%%%%%%%%%%%%%%%%%%%%%%%%%%%%%%%%%%%%%%%%%%%%%%%%%%%%%%%%%%%%
%%%%%%%%%%%%%%%%%%%%%%%%%%%%%%%%%%%%%%%%%%%%%%%%%%%%%%%%%%%%%%%%%%%%%%%%%%%

\section*{Appendix 2}

We list a class of solutions in presence of a perfect fluid. 
The energy-momentum
tensor is the same as in section 6, and the  matter is described by the
velocity field
\begin{displaymath}
u_i = e^{-\sigma}(1-k_0^{\ 2})^{-1/2}\ \delta^0_i.
\end{displaymath}
Thus, matter is described once more in the comoving frame, the 
velocity field fulfilling the condition $u_iu^i = e^{-2\sigma} + (k_iu^i)^2$
(cf. section 3). 

We consider solutions with a vector field in the form
\begin{equation}
k^i=H(1, \ -\frac{\boldsymbol x}{r}),
\end{equation}
$H$ beeing a constant. Switching to spherical coordinates, one finds
that (22) is a constant null-vector field. The field equations were 
solved in the spherically symmetric case by Dadhich \cite{Dad}, the
general solution beeing
\begin{equation}
e^{-\sigma}= c_1r^n+c_2 r^{2-n},
\end{equation}
$n$ beeing an integer and $H$ fulfilling the following condition:
\begin{equation}
(1+H)^2/(1-H) = 1+ 2n(n-2).
\end{equation}

As an 
example, we consider the case $n=-1$.  The solution then reads:
\begin{eqnarray*}
e^{-\sigma}&=& c_1r^{-1}+c_2r^3\\
\kappa\mu &=& \frac{3}{7}(c_2^2 r^4 +18 c_1c_2+c_1^2r^{-4})\\
\kappa p &=& \frac{1}{7}(9c_2^2r^4 - 38 c_1c_2 + c_1^2 r^{-4}).
\end{eqnarray*}
The pressure vanishes at $r_0^4 = (19 \pm \sqrt{352})c_1/9c_2$. This value 
 can thus be
interpretated as  the radius of the sphere of the matter distribution. Since we
still have two free parameters $c_1$ and $c_2$, it is possible to 
join the solution to the vacuum-solution (11) continously
at  $r=r_0$. To do this, we choose the constants in a way 
that at $r=r_0$, we have $\sigma =0$ as well as $H= \frac{a}{r_0}$. From
(24) we find $H = 1/2 (-9 \pm \sqrt{105})$, and thus for $r_0$, we have
the following equations:
\begin{eqnarray*}
c_1r_0^{-1}+c_2r_0^3 &=&1 \\
a/r_0 &=& (19 \pm \sqrt{352})\frac{c_1}{9c_2}.
\end{eqnarray*}

The full discussion can be found in the article mentioned above. 
\pagebreak

\end{document}